\documentclass[aps,prc,twocolumn,showpacs,superscriptaddress]{revtex4}

\usepackage{graphicx}
\usepackage{epstopdf}
\usepackage{multirow,dcolumn}

\begin{document}

\newcommand{\nnprime}{$n,n^\prime \gamma$}
\newcommand{\ntwon}{$n,2n\gamma$}
\newcommand{\nthreen}{$n,3n\gamma$}
\newcommand{\nxn}{$n,xn\gamma$}
\newcommand{\nx}{$n,x\gamma$}

\newcommand{\natpb}{$^{\textrm{nat}}$Pb}
\newcommand{\natge}{$^{\textrm{nat}}$Ge}
\newcommand{\eigpb}{$^{208}$Pb}
\newcommand{\sevpb}{$^{207}$Pb}
\newcommand{\sixpb}{$^{206}$Pb}
\newcommand{\fivpb}{$^{205}$Pb}
\newcommand{\foupb}{$^{204}$Pb}
\newcommand{\nonubb}  {$0 \nu \beta \beta$}
\newcommand{\twonubb} {$2 \nu \beta \beta$}
\newcommand{\gam}{$\gamma$}
\def\nuc#1#2{${}^{#1}$#2}
\def\mee{$\langle m_{\beta\beta} \rangle$}
\def\mnu{$\langle m_{\nu} \rangle$}
\def\ml{$m_{lightest}$}
\def\gnu{$\langle g_{\nu,\chi}\rangle$}
\def\mmod{$\| \langle m_{\beta\beta} \rangle \|$}
\def\mb{$\langle m_{\beta} \rangle$}
\def\BBz{$0 \nu \beta \beta$}
\def\BBm{$\beta\beta(0\nu,\chi)$}
\def\BBt{$2 \nu \beta \beta$}
\def\nonubb{$0 \nu \beta \beta$}
\def\twonubb{$2 \nu \beta \beta$}
\def\BB{$\beta\beta$}
\def\Mz{$M_{0\nu}$}
\def\Mt{$M_{2\nu}$}
\def\MzG{$M^{GT}_{0\nu}$}           
\def\MzF{$M^{F}_{0\nu}$}                
\def\MtG{$M^{GT}_{2\nu}$}           
\def\MtF{$M^{F}_{2\nu}$}                
\def\Tz{$T^{0\nu}_{1/2}$}
\def\Tt{$T^{2\nu}_{1/2}$}
\def\Tc{$T^{0\nu\,\chi}_{1/2}$}
\def\Rz{$\Gamma_{0\nu}$}            
\def\Rt{$\Gamma_{2\nu}$}            
\def\ms{$\delta m_{\rm sol}^{2}$}
\def\ma{$\delta m_{\rm atm}^{2}$}
\def\ts{$\theta_{\rm sol}$}
\def\ta{$\theta_{\rm atm}$}
\def\tot{$\theta_{13}$}
\def\gpp{$g_{pp}$}                  
\def\qval{$Q_{\beta\beta}$}                 
\def\MJ{{\sc Majorana}}             
\def\DEM{{\sc Demonstrator}}             
\def\be{\begin{equation}}
\def\ee{\end{equation}}
\def\cpRty{counts/ROI/t-y}
\def\onecpRty{1~count/ROI/t-y}
\def\fourcpRty{4~counts/ROI/t-y}
\def\ppc{P-PC}                          
\def\nsc{N-SC}                          
\def\dbd{double-$\beta$-decay}
\def\db{double $\beta$}


\title{Neutron inelastic scattering and reactions in natural Pb as a background in neutrinoless \dbd\ experiments}

\newcommand{\lanl}{Los Alamos National Laboratory, Los Alamos, NM 87545}
\newcommand{\usd}{Department of Earth Science and Physics, University of South Dakota, Vermillion, South Dakota 57069}

\author{V.~E.~Guiseppe} \email[Electronic address: ]{guiseppe@lanl.gov}\affiliation{\lanl}
\author{M. Devlin}\affiliation{\lanl}
\author{S.~R.~Elliott}\affiliation{\lanl}
\author{N. Fotiades}\affiliation{\lanl}
\author{A. Hime}\affiliation{\lanl}
\author{D.-M.~Mei}\affiliation{\usd}
\author{R.~O.~Nelson}\affiliation{\lanl}
\author{D.~V.~Perepelitsa}\affiliation{\lanl}
\date{\today}

\begin{abstract}
Inelastic neutron scattering and reactions on Pb isotopes can result in \gam\ rays near the signature end-point energy in a number of \BB\ isotopes. In particular, there are \gam-ray transitions in \nuc{206,207,208}{Pb} that might produce energy deposits at the \nuc{76}{Ge} \qval\ in Ge detectors used for \BBz\ searches. The levels that produce these \gam\ rays can be excited by (\nnprime) or (\nxn) reactions, but the cross sections are small and previously unmeasured. This work uses the pulsed neutron beam at the Los Alamos Neutron Science Center to directly measure reactions of interest to \BB-decay experiments. The cross section on \natpb\ to produce the 2041-keV \gam\ ray from \sixpb\ is measured to be 3.6  $\pm$0.7 (stat.) $\pm$ 0.3 (syst.) mb at $\approx$ 9.6 MeV. The cross section on \natpb\ to produce the 3061,3062-keV \gam\ rays from \sevpb\ and \eigpb\ is measured to be 3.9 $\pm$ 0.8 (stat.) $\pm$ 0.4 (syst.) mb at the same energy. We report cross sections or place upper limits on the cross sections for exciting some other levels in Pb that have transition energies corresponding to \qval\ in other \BB\ isotopes.
\end{abstract}


\maketitle
 

\section{Introduction}
\label{sec:Intro}



Neutrinoless \db\ decay (\BBz) plays a key role in understanding the neutrino's absolute mass scale and particle-antiparticle nature~\cite{Ell02, Ell04, Bar04, Avi05, eji05, avi08}. If this nuclear decay process exists, one would observe a monoenergetic line originating from a material containing an isotope subject to this decay mode. One such isotope that may undergo this decay is $^{76}$Ge.
Germanium-diode detectors fabricated from material enriched in $^{76}$Ge have established the best half-life
limits and the most restrictive constraints on the effective Majorana mass for the neutrino~\cite{aal02a,bau99}. One analysis~\cite{kla06} of the data in Ref.~\cite{bau99} claims evidence for the decay with a half-life of $2.23^{+0.44}_{-0.31} \times 10^{25}$ y.
Planned Ge-based \BBz\ experiments~\cite{gui08,sch05} will test this claim. Eventually, these future experiments target a sensitivity of $>$10$^{27}$ y or $\sim$1 event/ton-year to explore neutrino mass values near that indicated by the atmospheric neutrino oscillation results. 

The key to these experiments lies in the ability to reduce intrinsic radioactive
background to unprecedented levels and to adequately shield the detectors from external
 sources of radioactivity. Previous experiments' limiting backgrounds have been trace levels of natural decay chain isotopes within the detector and shielding components. The \gam-ray emissions from these isotopes can deposit energy in the Ge detectors producing a continuum, which may overwhelm the potential \BBz\ signal peak at 2039 keV. Great progress has been made identifying the location and origin of this contamination, and future efforts will substantially reduce this contribution to the background. The background level goal of 1 event/ton-year, however, is an ambitious factor of $\sim$400 improvement over the currently best achieved background level~\cite{bau99}. If the efforts to reduce the natural decay chain isotopes are successful, previously unimportant components of the background must be understood and eliminated. The work of Mei and Hime \cite{mei06} recognized that (\nnprime) reactions will become important for ton-scale \dbd\ experiments operated underground where the muon-induced neutrons extend to several GeV in energy.
 
 Reference~\cite{mei08} recognized that the specific \gam\ rays from Pb isotopes at 2041 and 3062 keV are particularly troublesome. The former is dangerously near the 2039.00$\pm$0.05-keV Q value for zero-neutrino \db\ decay in $^{76}$Ge and the latter can produce a double-escape peak line at 2041 keV. That article pointed out that the cross sections to produce these lines in \natpb\ were unmeasured and hence set to zero in the data bases of the simulation codes used to design and analyze \nonubb\ data. The article also attempted an initial estimate of the cross section, but made it clear that better measurements were required.  
 
 Previous authors have studied (\nxn) reactions in Pb usually using enriched samples to isolate the isotopic effects\cite{von94,mih06,mih08}. However, only~\cite{mei08} has attempted to report results for the small-cross-section transitions that produce the lines of interest for $^{76}$Ge \BBz.   
This article presents measurements of Pb(\nnprime) and Pb(\nxn)  production cross sections of the \gam\ rays near 2041 and 3062 keV from a broad-energy neutron beam.  A 2041(2)-keV \gam\ ray is produced by the 3743.7(7) keV level in \sixpb,  a 3062(10)-keV \gam\ ray from the 3633(2) level in \sevpb, and a 3060.82(2)-keV \gam\ ray from the 5675.37(2) level in \eigpb\ \cite{ensdf}. Although our work was motivated by neutron reaction considerations in materials that play important roles in the \MJ \cite{gui08} design, the results have wider utility because lead is used in numerous low-background experiments.

\section{Experiment}


Data were collected at the Los Alamos Neutron Science Center (LANSCE) Weapons Neutron Research (WNR) facility, which provides neutrons in the energy range from 0.2--800 MeV \cite{lis90}. The WNR facility was chosen for this work due to the broad neutron energy spectrum.
As the pulsed neutron beam strikes a Pb target, the outgoing \gam\ rays are detected by the GErmanium Array for Neutron Induced Excitations (GEANIE) spectrometer \cite{bec97}. GEANIE is located a distance of 20.34 m from the natural tungsten spallation target.


The neutron target at the center of GEANIE was five stacked foils of natural Pb (\natpb) angled 20$^{\circ}$ off the normal of the beam direction. Each foil measured nominally 5 cm $\times$ 5 cm in area and 0.475 mm in thickness. The GEANIE spectrometer consists of 26 HPGe detectors; 20 of which have BGO active shields. Of the 26 Ge detectors, 16 are coaxial geometry with a dynamic range up to 4 MeV and 10 are planar geometry detectors with a limited dynamic range of 1 MeV. Due to the high  \gam-ray energy region of interest for \nonubb, only the coaxial detectors were considered here. 
Many of the GEANIE coaxial detectors had reduced resolution, due either to neutron damage or other issues, and only the four detectors with the best energy and timing resolution were used in this analysis.

Some of the detectors have Pb collimators surrounding their face to select mainly radially directional \gam\ rays from the target. A majority of the data collected occured with the detectors in this configuration. This may present additional Pb for neutron interaction, but because this Pb was not in the path of the neutron beam, it is assumed that this contribution is negligible and therefore not considered in our analysis. We have taken data with a Cu target also. The experimental arrangement is the same but with the substitution of Cu for Pb in the target. These data show no peaks at the strongest transitions in Pb  and therefore neutron interactions in the Pb collimators do not appreciably contribute to our measured spectrum.

The pulsed neutron beam has the following timing structure. Macropulses, lasting 625 $\mu$s, occur at a rate of 40 Hz. Micropulses are spaced every 1.8 $\mu$s, during which the neutron energy is determined by the time of flight from the micropulse start. An in-beam fission chamber measures the neutron flux with $^{235}$U and $^{238}$U foils. There was one inch of borated polyethylene in the beam line to reduce the low energy wrap-around neutrons remaining after the start of the next micropulse. These wrap-around neutrons do not contribute to the production of any $\gamma$ rays for which the reaction threshold is above 650 keV. 
In addition to the \natpb\ data, there were source runs for calibration purposes. Data were collected from two radioactive sources, $^{152}$Eu and $^{226}$Ra, with the neutron beam shuttered. Analysis of the sources permit a full energy peak efficiency curve for the detector array.

\section{Analysis and Results}

\subsection{Analysis description}

The data collected from GEANIE are stored in an event mode file containing time and pulse-height information from each HPGe detector and the fission chamber. Data from each detector must be checked for integrity and aligned to adequately sum together with the others. The analysis permits gating the resulting Ge-detector energy spectra on neutron energy to achieve a neutron energy dependence on the \gam-ray production.

The \gam-ray detection efficiency of each detector was determined by constructing an efficiency curve based on the two sources counted. Using 22 lines in $^{152}$Eu and 29 in $^{226}$Ra, the known source activities and branching ratios, and the approximate run time, an efficiency curve was generated from a fit to the calculated efficiencies. The source run time is needed and used to calculate the number of disintegrations of the source during counting. Calculation of this quantity was limited to an estimation due to the use of an unstable, 60 Hz pulser. The scale of the absolute efficiency is corrected by the normalization discussed below.
A correction is  needed for attenuation of \gam\ rays in the target.  Based on the location of the four detectors used, a path length seen by each detector through the Pb target is determined. An energy-dependent attenuation correction is calculated by integrating the known exponential attenuation expression over the total path length through the Pb target. The energy-dependant attenuation correction is combined with the the detector specific full energy peak detection efficiency. The net full energy peak efficiency for the array is found as the sum of each detector's efficiency.
A MCNPX \cite{mcnpx} simulation code corrects for down-scattered neutrons in the target affecting the time-of-flight assignment of neutron energy \cite{nel01}.

With an array like GEANIE, observation of the angular dependance of the \gam\ rays is possible. However, with only four detectors used at four unique angles, this technique is not optimal. Instead, an integral \gam-ray production cross section is calculated based on the net \gam-ray yield of the four detectors.

Each \gam-ray event detected has timing information, which can be referenced to the start of a micropulse for time-of-flight neutron energy determination. The \gam-ray spectrum is gated into a number of individual spectra corresponding to a fixed time interval but increasing neutron-energy width. The \gam-ray lines of interest are fitted via the ``gf3" component of the RADWARE package \cite{radware} to obtain \gam-ray yields as a function of neutron energy. 

The neutron spectra measured by the fission chamber are also gated into bins relative to the micropulse start signal via the time-of-flight technique used to bin the \gam-ray spectrum. The time-of-flight calculation accounts for the fission chamber being upstream of the target.
The gated fission chamber pulse height spectra are analyzed using known neutron-induced fission cross sections of $^{238}$U to determine neutron yield as function of neutron energy.

The angle-integrated \gam-ray cross section [$\sigma_\gamma(E_n)$] can be calculated using
\begin{equation}
\sigma_\gamma(E_n)=\frac{I_\gamma}{T_\gamma \, \epsilon_\gamma} \frac{T_\Phi}{\Phi}
\frac{1+\alpha}{t}\,N
\label{eq:cs}
\end{equation}
where $I_\gamma$ is the \gam-ray yield (counts/MeV), $\Phi$ is the neutron yield (neutrons/MeV), $\epsilon_\gamma$ is the absolute \gam-ray detection efficiency, $t$ is the target aerial density (atoms/barn), $\alpha$ is the internal conversion coefficient,  and $T_\gamma$ and $T_\Phi$ are the fractional live times of the HPGe detectors and fission chamber, respectively. The live-time fractions are nearly identical due to similar electronics in all channels of GEANIE. Because the yield of neutrons and \gam\ rays show up as a ratio, the time interval over which data were collected is not needed in Eq. (\ref{eq:cs}).
$N$ is a normalization factor based on a known cross section. 

If the normalization factor is calculated from the same data set, the energy-independent factors cancel leaving
\begin{equation}
\sigma_\gamma(E_n)=\frac{I_\gamma}{\epsilon_\gamma} \frac{1+\alpha}{\Phi}
N^{\textrm {eff}}
\end{equation}
An effective normalization factor, $N^{\textrm{eff}}$, is defined as
\begin{equation}
N^{\textrm{eff}}= \sigma_{K}\,\frac{\epsilon_\gamma^K}{I_\gamma^K} \frac{\Phi^K}{1+\alpha^K}
\end{equation}
where $\sigma_{K}$ is a known cross section in \natpb\ corresponding to a given transition labeled $K$. Neglecting the energy-independent contributions to the cross section allows a reduction in the uncertainty budget.

The prominent lines in the major Pb isotopes have been studied in detail \cite{von94,mih06,mih08} and serve as a convenient reference and known transition for normalization. The recent study \cite{mih06,mih08} of Pb isotopes showed good agreement between past experimental results and that of the TALYS \cite{kon05} nuclear reaction code. The TALYS code ran default settings with proper treatment of long-lived isomers. The pulsed structure of the LANSCE beam and the time-of-flight gating allows the detection of prompt decays while long-lived isomers are a subtracted background. 

Three of the strongest lines in \natpb\ are analyzed and compared against the TALYS result for normalization. The three first excited state \gam-ray transitions are the 2614.5-keV \gam\ ray in \eigpb, the 569.7-keV \gam\ ray from \sevpb, and the 803.1-keV \gam\ ray from \sixpb. These three major lines were analyzed using a 15-ns time-of-flight neutron binning, which is the resolution limit of the HPGe detectors. Gamma-ray production cross sections are calculated using TALYS to represent the production in \natpb\ by summing and weighting by isotopic abundance. Therefore the 803-keV \gam\ ray can be produced through the following reaction channels: \sixpb(\nnprime)\sixpb, \sevpb(\ntwon)\sixpb, and \eigpb(\nthreen)\sixpb. Similarly, the 570-keV \gam\ ray can be produced through the two channels: \sevpb(\nnprime)\sevpb\ and \eigpb(\ntwon)\sevpb. The 2615-keV \gam\ ray only results from \eigpb(\nnprime)\eigpb. To obtain the normalization factor, the output of TALYS and experimental results were compared over the dominant neutron range of 4--7 MeV where the production rates for these lines are strongest. 
The normalization factor resulting from this analysis is estimated to be $N=1.22 \pm 0.03$ (2.5\%). 

\subsection{Cross sections}
\label{sec:cs}

\begin{figure}[t]
\includegraphics[width=8.6cm]{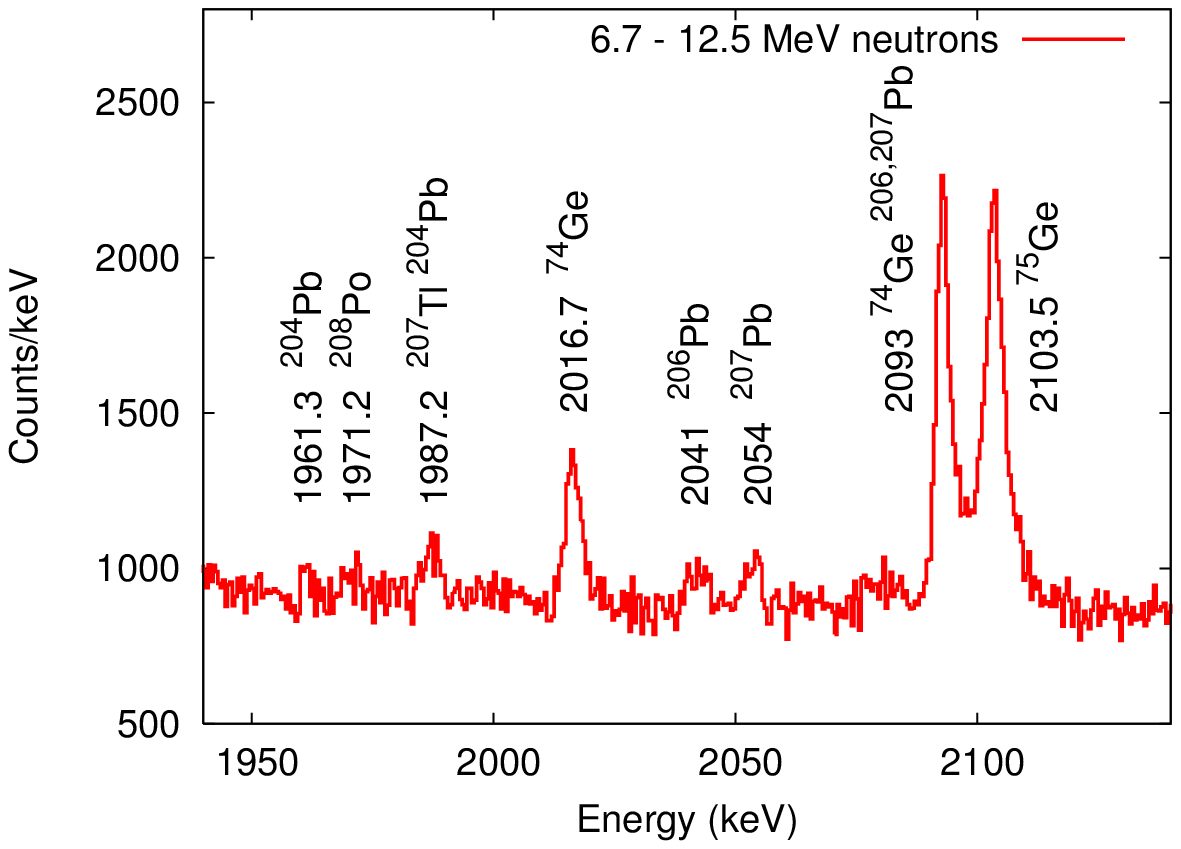}\\
\includegraphics[width=8.6cm]{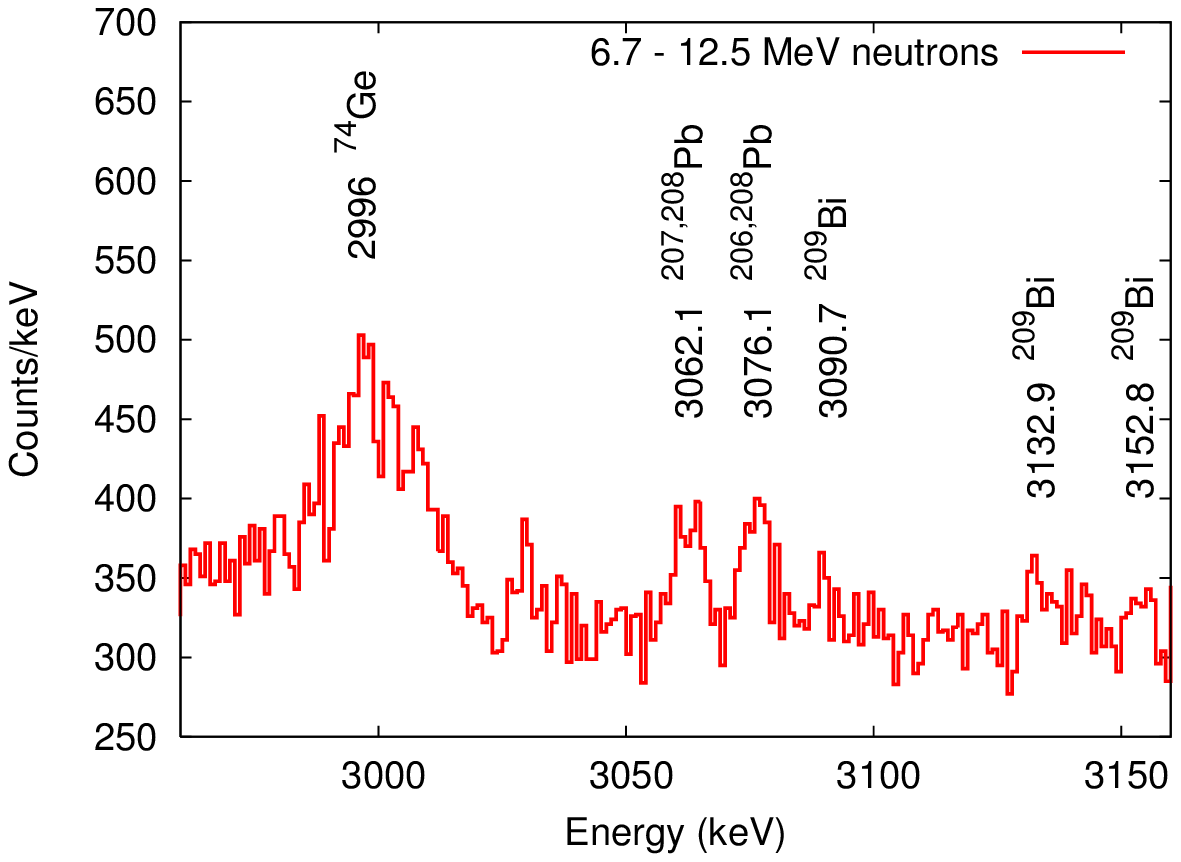}
\caption{(Color online) The \gam-ray spectra showing the (a) 2041-keV \gam-ray line and the (b) 3062-keV \gam-ray line (6.7-12.5 MeV neutrons).  \label{fig:mj-spec}}
\end{figure}

The two lines of most interest to Ge-based \nonubb\ experiments using Pb shielding are weak and were observable in the data only by summing spectra over a large range of neutron energy (Fig. \ref{fig:mj-spec}). The \gam-ray spectra were gated by a larger bin width of 150 ns, which resulted in nine bins covering the energy range of 1--200 MeV. A peak assigned    2041 keV ranged $\pm$ 1.4 keV and is  most dominant in the 4--30 MeV neutron range. The second peak assigned 3062 keV ranged $\pm$0.7 keV and is most dominant in the range from 4--13 MeV neutron range.

The presence of a 3062-keV \gam\ ray indicates the possibility of a double escape peak near 2040 keV  contributing to the peak assigned to a 2041-keV \gam\ ray. The double escape peak of the 3062-keV \gam\ ray was calculated from the ratio of the observed 2614.5-keV \gam\ ray double escape peak and full energy peak and corrected for the relative efficiency of a full energy peak and the cross section of pair production \cite{gud80,eul74,cha67} between a 3062- and a 2614.5-keV \gam\ ray. The probability of electron-positron absorption and annihilation \gam\ double escape were assumed to be the same for the  3062- and 2614.5-keV \gam\ rays. 

The feature near 2041 keV appears to be a blend, although the low intensity of the peak results in a lack of certainty. If one assumes the feature includes two peaks, one gets a better fit with roughly two equal peaks at 2040 and 2044 keV. Alternatively, if one assumes a lone peak, the width is wider than nearby peaks and there is a neutron-energy dependence of the centroid of $\pm$1.4 keV. A candidate transition for the additional peak is the 2046.1(1)-keV \gam\ ray from the 4111.34(8)-keV level in $^{204}$Pb. The low isotopic abundance of $^{204}$Pb makes this a dubious but possible assignment. A two-peak fit indicates that the 2040-keV and 2044-keV transitions both show significant relative strength at a neutron energy between 5 and 13 MeV. 
This large range of neutron energies suggests that both (\nnprime) and (\nxn) reactions are feeding the possible feature at 2044 keV and if not due to $^{204}$Pb, it must still be fed by other \natpb\ reactions.

We have also tried to identify an isotope with the apparatus, but not the target, that could result in $\gamma$ rays between 2035 and 2050 keV. The concern is that scattered neutrons might interact with materials that don't reside within the neutron beam. However, we have also taken data with a Cu target, that has the same geometry as the Pb. There is no line near 2040 or 2044 keV in those data, and therefore possible transitions that might arise from materials surrounding the target, but not within the neutron beam (such as the Ge detectors, BGO detectors, or the Pb collimators), cannot be the source of a $\gamma$ ray.  Hence, although there is some uncertainty in assigning the specific process producing the $\gamma$ rays contributing to the feature near 2041 keV, the evidence indicates they do originate from the Pb. Therefore we fit the feature as a single peak and report an effective cross section for producing $\gamma$ rays at 2041 keV from neutron interactions in natural Pb.

The feature near 3062 keV  is expected to be a blend with contributions from a 3062(10)-keV and 3060.82(2)-keV \gam\ rays. Due to the low intensity of the peak and the proximity of the two expected contributions of the doublet, the feature was fit as a single peak.
For those bins where a peak was not observed, a sensitivity limit was measured to provide an upper limit on the cross section. The internal conversion coefficients for these three transitions are not known experimentally and are assumed to be negligible.

\begin{figure}
\includegraphics[width=8.6cm]{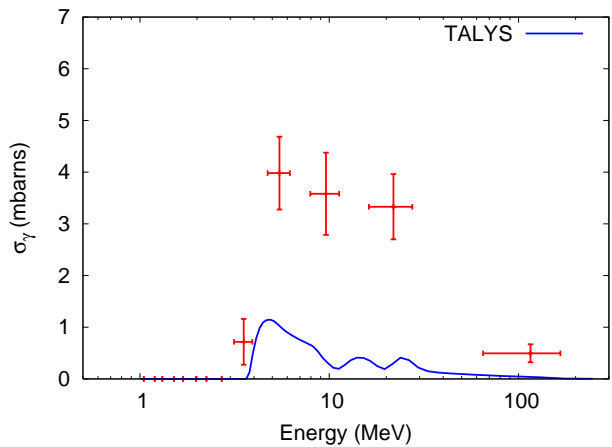}
\caption{(Color online) The measured \gam-ray production cross section for the 2041-keV \gam\ ray in \sixpb. The solid line is the predicted cross section calculated by the TALYS nuclear reaction code for the expected \gam\ ray in \natpb. \label{fig:2040}}
\end{figure}

\begin{figure}
\includegraphics[width=8.6cm]{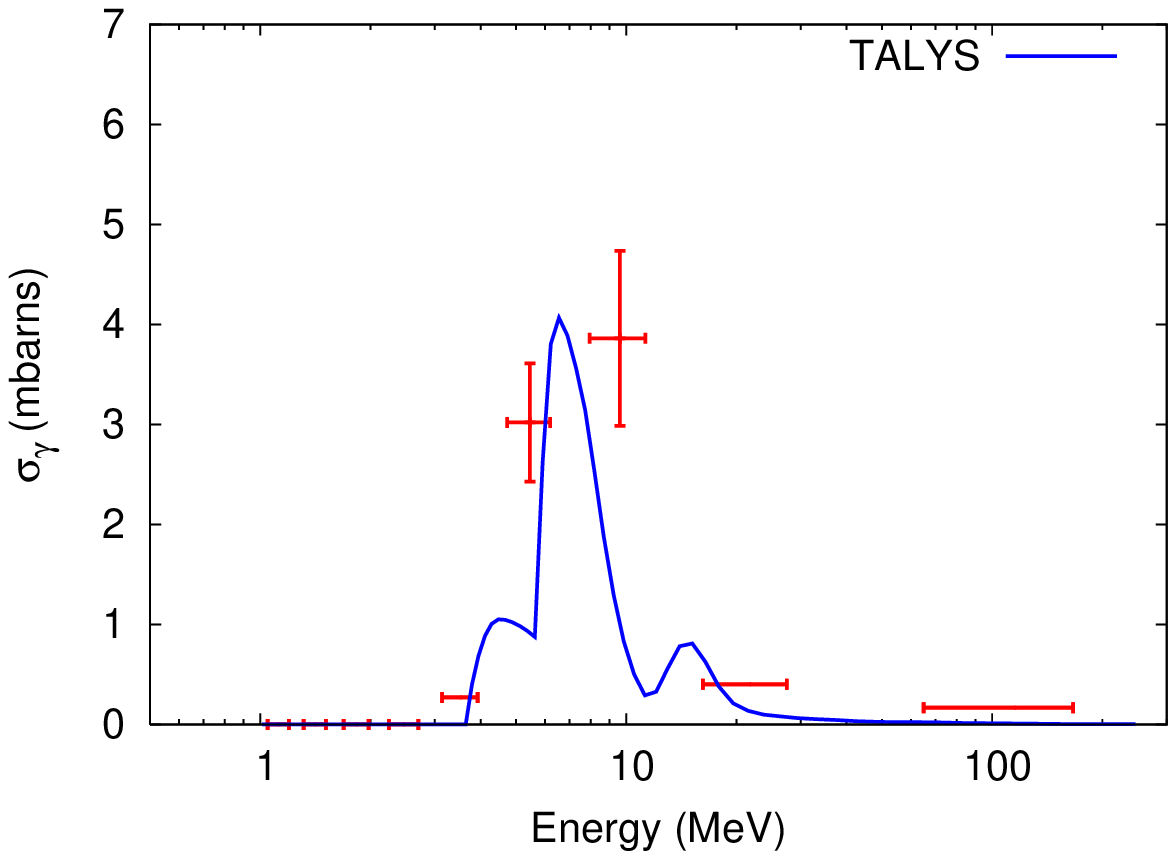}
\caption{(Color online) The measured \gam-ray production cross section for the 3061-/3062-keV \gam\ rays in \nuc{207,208}{Pb}. The solid line is the predicted cross section calculated by the TALYS nuclear reaction code for the expected \gam\ rays in \natpb. \label{fig:3062}}
\end{figure}

\begin{table*}
\caption{The summary of the cross sections as a function of energy for the 2041- and 3061-/3062-keV transitions. }
\label{tab:results}
\begin{tabular}{lcc}

& \multicolumn{2}{c}{Cross section (mb)}\\
\cline{2-3}
Neutron energy     & \natpb(\nxn)\sixpb                                                  &  \natpb(\nxn)\nuc{207,208}{Pb}\\
(MeV)                      &  2041 keV                                                                &  3061,3062 keV\\
\hline\hline
2.87--4.20                & 0.72  $\pm$ 0.44(stat.) $\pm$ 0.07 (syst.)        & $<$0.3\\
4.20--6.72                 & 4.0  $\pm$ 0.6 (stat.) $\pm$ 0.4 (syst.)             & 3.0 $\pm$ 0.5 (stat.) $\pm$ 0.3 (syst.) \\
6.72--12.50             & 3.6  $\pm$ 0.7 (stat.) $\pm$ 0.3 (syst.)              & 3.9 $\pm$ 0.8 (stat.) $\pm$ 0.4 (syst.)\\
12.50--31.15            & 3.3 $\pm$ 0.6 (stat.) $\pm$ 0.3 (syst.)             & $<$0.4\\
31.15--200               & 0.50  $\pm$ 0.17 (stat.) $\pm$ 0.05 (syst.)        & $<$0.2\\

\hline 
\end{tabular}
\end{table*}

The cross sections for the 2041- and 3062-keV \gam-ray lines are shown in Figs. \ref{fig:2040}--\ref{fig:3062} and tabulated in Table \ref{tab:results}. The results are 3--4 mb for both the 2041- and 3062-keV lines  for neutron energies 4--13 MeV. The figures show a prediction from the TALYS nuclear reaction code. For the line assigned 2041 keV, the TALYS calculation includes the 2041(2)-keV \gam\ ray from the \sixpb(\nnprime)\sixpb, \sevpb(\ntwon)\sixpb, and \eigpb(\nthreen)\sixpb\ reactions  in \natpb. For the line assigned 3062 keV, the TALYS calculation includes the 3062(10)-keV \gam\ ray from the \sevpb(\nnprime)\sevpb\ and \eigpb(\ntwon)\sevpb\ reactions and the 3060.82(2)-keV \gam\ ray  from the \eigpb(\nnprime)\eigpb\ reaction in \natpb. The TALYS code used primarily default input settings; the maximum number of included discrete levels considered in Hauser-Feshbach \cite{hau52} decay and the gamma-ray cascade was increased to 150 levels from the default value to excite the levels of interest. Therefore, the highest included level was 4901(4) keV in \sixpb, 6090(7) keV in \sevpb, and 6420.2(14) in \eigpb.

The experimental result for the 3062-keV \gam-ray is consistent with that predicted from the TALYS calculation.  However, the 2041-keV \gam-ray result exceeds the prediction of TALYS. One possible explanation is the presence of another source of \gam\ rays near 2041 keV from materials in the experiment or including other reactions in the Pb target. However, as mentioned previously, a run with a Cu target showed no peak in the spectrum near 2041 keV and we suspect the peak we show is due to the Pb target.
The TALYS simulation code uses the Hauser-Feshbach technique to estimate inelastic neutron scattering cross sections. Our work measures cross sections for producing $\gamma$ rays from high-energy excited nuclear levels whose spin-parity and \gam-ray branching ratios are often not known, which causes the Hauser-Feshbach model calculation to be uncertain. Hence, although we performed TALYS calculations to estimate the cross sections and anticipated rates in the experiment,  we do not consider the calculations  as a reliable cross-check of our experimental results.

\subsection{Measurement Uncertainty}

The systematic and statistical uncertainties encountered are listed in Table \ref{tab:error}.  The normalization correction uncertainty is estimated from the agreement of normalization factors obtained from the three prominent lines compared against known reference cross sections. The three known reference transitions have an experimental uncertainty of less than $\sim$7\% in the energy range 3--10 MeV and so we include a 7\% systematic uncertainty. The efficiency curve uncertainty derives from the error in the fit to the source data and the error in the attenuation correction, which is dominated by the uncertainty in the parameterized mass attenuation coefficient \cite{gud80} and the tolerance of the target thickness. The statistical errors in the flux and yield measurements are based on the counts in the fission chamber and \gam-ray detectors, respectively. The uncertainty of the neutron energy is based on the timing resolution of the HPGe detectors. The statistics of the weak lines of interest are the dominant source of uncertainty.

\begin{table}[t]
\begin{center}

\caption{A listing of the systematic and statistical uncertainties in the cross section measurement. \label{tab:error}}
\begin{tabular}{lcc}

& \multicolumn{2}{c}{Systematic Uncertainty }\\
\cline{2-3}
                         & 2041-keV  & 3061-/3062-keV  \\
                         & \gam\ ray   & \gam\ ray \\
\hline\hline
Normalization & 2.5\%            & 2.5\% \\
Reference transitions	& 7\% & 7\% \\
Efficiency        & 1.22\%       & 1.24\% \\
Angular correction & 6\% & 6\%\\

\hline
\end{tabular}
\end{center}

\begin{tabular}{lccc}

& \multicolumn{3}{c}{Statistical Uncertainty }\\
\cline{2-4}
Neutron energy~~~~~~& Neutron flux & 2041-keV & 3061,3062 keV \\
(MeV)              &                                & \gam-ray yield & \gam-ray yield \\
\hline\hline
1.00--1.24        & 5.1\%                    &                            &\\
 1.24--1.58       & 2.2\%                     &                            &\\
1.58--2.08        & 1.0\%                     &                            & \\
 2.08--2.87       & 0.92\%                     &                            & \\
 2.87--4.20       & 0.87\%                     &      61\%           & \\
 4.20--6.72       & 0.76\%                     &       15\%           & 17\% \\
 6.72--12.50     & 0.62\%                     &      20\%         & 21\% \\
 12.50--31.15  & 0.60\%                     &         17\%       & \\
  31.15--200    & 0.49\%                     &         34\%        & \\

\hline 
\end{tabular}
\end{table}

Generally, the angle-integrated \gam-ray cross section is computed from an integral of the differential cross section.  The \gam-ray differential cross section can be expanded in a sum of Legendre polynomials and, excluding contribution from cascades, can be predicted for transition between some states near threshold \cite{she66}. Higher energies and contributions from cascades diminish anisotropy in the angular distribution. In this experiment, the \gam-ray yield was the sum of the individual yields of four detectors and assumes an isotropic distribution. Ref. \cite{she66, mih06} define anisotropic distributions for the $E2$, $E3$, and $M1+E2$ multipolarities. If these three anisotropic distributions are evaluated at the locations of the four detectors used ($\cos \theta$ = 0.1818, 0.2267, 0.5492, 0.8745), there is no more than a 6\% deviation from isotropic assumption. Further, Ref. \cite{fot01} tabulates angular distribution corrections for a \nuc{238}{U} target when assuming an isotropic distribution and finds corrections mainly within 5\%. We include a 6\% systematic uncertainty to account for any \gam-ray angular distribution.

\section{Discussion and Conclusion}
The cross sections of interest to Ge \BBz\ are tabulated in Table \ref{tab:results}.
In Ref.~\cite{mei08} an estimate for the cross section of the $^{207}$Pb($n,n^\prime$ 3062-keV \gam\ ray)$^{207}$Pb was given as 75 mb with an
uncertainty of about 20\%. For \natpb\ this would indicate a cross section for comparison of $\approx$16 mb at an average neutron energy of $\approx$4.5 MeV. Considering the 
caveats discussed in the previous work with regards to the presence of Cl, the fact that the cross section is above the value reported in
this work is considered to be consistent. Reference~\cite{mei08} provided no value for the cross section for the production of the 2041-keV \gam\ ray.

\begin{table*}[t]

\caption{A list of frequently studied isotopes for \BB\ and their $Q$ values\cite{aud03,rah07,red07}. In columns 3--5 are indicated Pb isotopes
with levels that emit \gam\ rays near the energies of the $Q$ value, the $Q$ value + 511 keV (SEP, single escape peak), and the $Q$ value + 1022 keV (DEP, double escape peak), respectively. Also in columns 3--5 are the cross-section values or upper limits for exciting the transitions in \natpb\ with neutrons of energy 6.72--12.50 MeV. Where the cross section is listed as NA, we were unable to place a limit due to the \gam-ray energy being outside the range of the detection system. The line arguably present at the SEP for $^{116}$Cd is due to a transition we are unable to identify. }
\label{tab:isotopes}
\begin{tabular}{|c|c|c|c|c|}

\hline  
\BB\ isotope         & \qval\ (keV)                               &  \gam\ ray                                              		& SEP                                         	 		&  DEP          \\
\hline\hline
\nuc{76}{Ge}       &   $2039.00 \pm 0.05$              &    \sixpb\   $\sigma = 3.6 \pm 0.8$ mb           	&                                                    			& \nuc{207,208}{Pb}      $\sigma = 3.9 \pm 0.9$ mb      \\
\nuc{82}{Se}       &   $2995.5 \pm 1.9$                   &                                                              		 &                                                    			&    \nuc{208}{Pb} $\sigma$ NA         \\
\nuc{100}{Mo}    &   $3034.40 \pm 0.17$               &   		\nuc{208}{Pb}	$\sigma <0.4 $ mb 	&  \sixpb\ $\sigma = 2.7 \pm 0.6$ mb    	&      \nuc{206}{Pb} $\sigma$ NA    \\
\nuc{116}{Cd}    &   $2809 \pm 4$                          &                                                               		&     $\sigma = 0.69 \pm 0.49$ mb       &                      \\
\nuc{130}{Te}     &   $2530.3 \pm 2.0$                   &                                                               		& 	\nuc{208}{Pb} 	$\sigma <0.4 $ mb	&                \\
\nuc{136}{Xe}     &   $2457.83 \pm 0.37$              &  \nuc{206,208}{Pb}       $\sigma <0.3 $ mb        		&                                                    			&                    \\
\nuc{150}{Nd}     &   $3367.7 \pm 2.2$                  &                                                              		&                                                    			& \sevpb\        $\sigma$ NA       \\
\hline 
\end{tabular}
\end{table*}

Table~\ref{tab:isotopes} lists a number of frequently used isotopes for \BB\ and their $Q$ values. In addition, the table lists Pb isotopes that have transitions near three critical energies for each isotope. A \gam\ ray with an energy similar to that of the $Q$ value can produce a background line that might mimic \BBz\ in an experiment that uses a Pb as a shield or for some other apparatus component. In addition, however, \gam\ rays that pair produce in a detector but have either one (single escape peak) or both (double escape peak) annihilation \gam\ rays escape the detector may also produce a line feature at the \BBz\ end point for a given initial \gam-ray energy. Such \gam\ rays would be 511 keV and 1022 keV more energetic than the $Q$ value. Also in Table~\ref{tab:isotopes}, we provide cross sections or upper limits on the cross sections to excite the levels that produce these \gam\ rays in \natpb.

We have measured the cross sections for the production of \gam\ rays of interest to Pb-shielded Ge \BBz\ experiments relative to previously measured strong transistions and the results are presented in Table~\ref{tab:results}. The cross sections are small (few mb) near 5--10 MeV and are below our sensitivity by about 40 MeV. The cross sections can be folded with the underground neutron flux to estimate background rates for such experiments. Although, the rates will likely be very low, the overall background must be extremely low to have the required sensitivity to \BBz. 
\begin{acknowledgments}
This work was supported in part by Laboratory Directed Research and
Development at Los Alamos National Laboratory and National Science Foundation grant 0758120. This work benefited from the use of the Los Alamos Neutron Science Center, funded
 by the U.S. Department of Energy under contract DE-AC52-06NA25396. We thank Toshihiko Kawano for discussions related to the use of TALYS.
\end{acknowledgments}


\bibliography{neutron}

\end{document}